%% file: 6-5-2.tex
\documentclass[11pt,twoside]{article}
\usepackage{atmp04}
\copyrightnotice{2002}{6}{827}{846} 

\usepackage{cite,latexsym,amsfonts,exscale,epsfig} 

\let\mathbbm\mathbb

\newcommand{\color}[2][c]{}

\theoremstyle{definition}
\newtheorem{theorem}{Theorem}[section]
\newtheorem{lemma}[theorem]{Lemma}
\newtheorem{corollary}[theorem]{Corollary}
\newtheorem{proposition}[theorem]{Proposition}
\newtheorem{definition}[theorem]{Definition}
\newtheorem{remark}[theorem]{Remark}

\numberwithin{equation}{section}

\def\acknowledgments{\section*{Acknowledgments}}%

\def\nn{\notag}

\def\emph#1{{\sl #1\/}}

\def\SO{{SO}}

\def\SU{{SU}}

\def\Spin{{Spin}}

\def\C{{\mathbbm C}}

\def\ie{{\sl i.e.\/}}

\def\1{\mathbf{1}}
\def\Calg{C_{\rm alg}}%
\def\H{\sym{H}}%
\def\L{\sym{L}}%

\def\sym#1{{\mathcal #1}}

\let\phi=\varphi
\let\theta=\vartheta
\let\epsilon=\varepsilon

\def\tr{\mathop{\rm tr}\nolimits}
\def\ev{\mathop{\rm ev}\nolimits}
\def\coev{\mathop{\rm coev}\nolimits}

\def\dim{\mathop{\rm dim}\nolimits}

\def\Hom{\mathop{\rm Hom}\nolimits}
\def\Aut{\mathop{\rm Aut}\nolimits}

\def\mycaption#1#2{%
  \begin{quote}
  \caption{\label{#1}#2}
  \end{quote}}

\newenvironment{myenumerate}{%
  \begin{enumerate}
  \setlength{\partopsep}{0pt}
  \setlength{\parskip}{0pt}}{\end{enumerate}}
 
\begin{document}
%

\title{Positivity of relativistic spin network evaluations}
\arxurl{gr-qc/0211106}
\author{Hendryk Pfeiffer}
\address{Perimeter Institute for Theoretical Physics,\\
         35 King Street N, Waterloo, Ontario, N2J 2W9, Canada\\
         and\\
         Emmanuel College, St Andrew's Street,\\ 
         Cambridge CB2 3AP, United Kingdom}
\addressemail{hpfeiffer@perimeterinstitute.ca}

\markboth{\it Positivity of relativistic spin network evaluations}{\it Hendryk Pfeiffer}


%
\begin{abstract}
%

Let $G$ be a compact Lie group. Using suitable normalization
conventions, we show that the evaluation of $G\times G$-symmetric spin
networks is non-negative whenever the edges are labeled by
representations of the form $V\otimes V^\ast$ where $V$ is a
representation of $G$, and the intertwiners are generalizations of the
Barrett--Crane intertwiner. This includes in particular the
relativistic spin networks with symmetry group $\Spin(4)$ or $\SO(4)$
on a large class of graphs, not restricted to the graph underlying the
$10j$-symbol. We also present a counterexample, using the finite group
$S_3$, to the stronger conjecture that all spin network evaluations
are non-negative as long as they can be written using only group
integrations and index contractions. This counterexample applies in
particular to the product of five $6j$-symbols which appears in the
spin foam model of the $S_3$-symmetric $BF$-theory on the two-complex
dual to a triangulation of the sphere $S^3$ using five tetrahedra. We
show that this product is negative real for a particular assignment of
representations to the edges.

\end{abstract}

\cutpage

%
\section{Introduction}
%

Spin networks were invented by Penrose as a tool to describe the
quantum geometry of space-time~\cite{Pe71}. They feature as the
kinematical states in non-perturbative Quantum Gravity~\cite{RoSm95}
and play a central role in Spin Foam Models~\cite{Ba98a} of Quantum
Gravity whose amplitudes are calculated by evaluating spin networks.

A spin network is a graph whose edges are labeled by representations
of a suitable symmetry group $G$ and whose vertices are labeled by
compatible intertwiners ($G$-morphisms). A spin network is evaluated
by writing down tensors for the intertwiners at the vertices and by
contracting their indices as prescribed by the edges. 

In the Barrett--Crane spin foam model~\cite{BaCr98} of Riemannian
Quantum Gravity, the amplitudes of the path integral are defined in
terms of a special type of spin networks. These are called the
\emph{relativistic spin networks}. Their symmetry group is
$\Spin(4)\cong\SU(2)\times\SU(2)$ or $\SO(4)$, their edges are labeled
by \emph{balanced representations}, \ie\ representations of the form
$V\otimes V$ where $V$ denotes a finite-dimensional irreducible
representation of $\SU(2)$, and their vertices are labeled by a
special intertwiner, known as the \emph{Barrett--Crane
intertwiner}. The four-simplex amplitude of the Barrett--Crane model
is given by a particular relativistic spin network, the
\emph{relativistic $10j$-symbol}, whose underlying graph is the
complete graph of $5$ vertices.

In the course of the first explicit computations of relativistic
$10j$-symbols \cite{ChEg02}, it was observed that they always evaluate
to non-negative real numbers, up to some signs which cancel when one
calculates the product of $10j$-symbols over all four-simplices of a
closed manifold~\cite{BaCh02}.

In the present article, we generalize the result of~\cite{BaCh02} to
relativistic spin networks on a large class of graphs\footnote{Any
subgraph of a complete graph.}. While~\cite{BaCh02} has established
the positivity of the Barrett--Crane amplitudes for the case of
$10j$-symbols, \ie\ for five-valent vertices and therefore for a model
which is defined on the two-complex dual to a triangulation, our
generalization extends this result to the Barrett--Crane model defined
on generic two-complexes. In addition, we present a formulation in
which unnecessary signs are avoided right from the beginning and which
allows us to generalize the result to $G\times G$-symmetric spin
networks whose edges are labeled by representations of the form
$V\otimes V^\ast$ where $V$ denotes a representation of $G$, and whose
intertwiners are certain generalizations of the Barrett--Crane
intertwiner.

The key idea of our proof is to use a canonical description for the
intertwiners in terms of their integral presentation. A main technical
problem in the study of spin networks is that one needs good
conventions in order to define the intertwiners, \ie\ some
Clebsch--Gordan coefficients, at the vertices. For a single
intertwiner, however, there is no canonical definition known. Writing
down Clebsch--Gordan coefficients rather requires the choice of bases
for the representation spaces, and the resulting expressions do depend
on these choices.

A typical example is a three-valent vertex of an $\SU(2)$-spin network
whose edges are labeled by irreducible representations. Let
$V_j,V_k,V_\ell$ where $j,k,\ell=0,\frac{1}{2},1,\ldots$ denote the
irreducible representations of $\dim V_j=2j+1$. Then the
dimension of the space of compatible intertwiners is
\begin{equation}
\label{eq_su2trivalent}
  \dim\Hom_{\SU(2)}(V_j\otimes V_k\otimes V_\ell,\C)=\left\{
    \begin{matrix}
      1,&\mbox{if}\quad \ell=|j-k|,|j-k|+1,\ldots,j+k,\\
      0,&\mbox{otherwise}.
    \end{matrix}\right.
\end{equation}
The space of intertwiners is at most one-dimensional, but this still
does not fix the intertwiner. Given any choice of normalization, there
will still be signs appearing when one exchanges two of the three
tensor factors. This can be seen most easily in the special case
$j=k=\ell=1$ in which the one-dimensional trivial representation is
contained in the totally antisymmetric subspace of $V_1\otimes
V_1\otimes V_1$. The main difficulty of the positivity proof are these
signs which one has to keep track of. The positivity proof that has
been found for products of $10j$-symbols~\cite{BaCh02} already
indicates that it is necessary to consider pairs of these
intertwiners, chosen carefully so that these signs cancel.

The strategy of the present article is to concentrate on canonical
objects which can be defined without any choices\footnote{Robert Oeckl
conjectured that one can produce a large class of positive spin
networks following this idea.}. Already in the study of the duality
transformation for non-Abelian lattice gauge
theory~\cite{OePf01,Pf01}, it was noticed that an important role is
played by a canonically defined object, the intertwiner arising from
the integration over the symmetry group $G$ acting on a tensor product
of representations, equation~\eqref{eq_haarinter} below, which gives
rise to pairs of intertwiners, called $P^{(j)}$
in~\eqref{eq_haarinterproj} below, whose \emph{relative} normalization
is canonically fixed. This leads automatically to the desired
cancellation of signs.

The evaluated spin networks whose positivity we wish to prove, take in
general values in $\C$ and can be written as traces of suitable linear
maps. At the technical level, the key idea of the proof is to defer
the calculation of the traces to a later stage and to study the linear
maps instead. In fact, many of these linear maps are actually
\emph{positive}, \ie\ it is not only their trace that is positive, but
rather the individual summands of the trace. By making these stronger
conditions explicit in the calculations, we are able to construct an
infinite family of positive linear maps by induction. Their traces
will then provide the evaluated spin networks and establish their
positivity. The proof is elementary and purely algebraic except for
the input that orthogonal projectors are positive.

Going beyond the study of relativistic spin networks, it is tempting
to conjecture that all spin networks that can be written down using
only group integrations and index contractions, are positive. We show
by counterexample that this stronger conjecture is not true.

The present article is organized as follows. In
Section~\ref{sect_prelim}, we review some properties of positive
operators and of finite-dimensional representations of compact Lie
groups. We also introduce a convenient diagrammatic language. We
prove the positivity of relativistic spin networks in
Section~\ref{sect_positive} and present counterexamples to the
stronger conjecture in
Section~\ref{sect_counterexample}. Section~\ref{sect_conclusion}
contains some concluding remarks.

%
\section{Mathematical background}
%
\label{sect_prelim}

\subsection{Positive linear operators}

Basic facts about positive linear maps can be found in many
textbooks. We need only properties that hold in (possibly
infinite-dimensional) Hilbert spaces and refer the reader
to~\cite{ReSi72} for more details and for the proofs of the following
results.

\begin{definition}
Let $\H$ be a Hilbert space with scalar product
$\left<\cdot,\cdot\right>$. We denote the set of bounded linear
operators on $\H$ by $\L(\H)$. An operator $A\in\L(\H)$ is called
\emph{positive} if $\left<v,Av\right>\geq 0$ for all $v\in\H$. In this
case, we write $A\geq 0$.
\end{definition}

For any operator $A\in\L(\H)$, we have $A^\dagger A\geq 0$. Positive
linear operators have the following properties.
\begin{lemma}
Let $\H$ be a Hilbert space and $A\in\L(\H)$ be positive.
\begin{myenumerate}
\item
  $A$ is self-adjoint, $A=A^\dagger$,
\item
  There exists a unique positive $B\in\L(\H)$ such that $B^2=A$,
\item
  Any operator $X\in\L(\H)$ for which $[A,X]=0$, also satisfies
  $[B,X]=0$.
\end{myenumerate}
\end{lemma}

In this article, we will construct our positive operators from
orthogonal projectors.
\begin{definition}
Let $\H$ be a Hilbert space. A linear operator $P\in\L(\H)$ is called
a \emph{projector} if $P^2=P$. A projector $P$ is called
\emph{orthogonal} if $P^\dagger=P$.
\end{definition}

We can then construct positive operators as follows.
\begin{lemma}
\label{lemma_positivity}
Let $\H,\H_1,\H_2$ be Hilbert spaces.
\begin{myenumerate}
\item
  Any orthogonal projector $P\in\L(\H)$ is positive,
\item
  If $A\in\L(\H_1)$ is positive, so is
  $A\otimes\1_{\H_2}\colon\H_1\otimes\H_2\to\H_1\otimes\H_2$, where
  $\1_{\H_2}$ denotes the identity on $\H_2$,
\item
  If $D\colon\H_1\otimes\H_2\to\H_1\otimes\H_2$ is positive and the
  partial trace\\ $\tr_{\H_2}(D)\colon\H_1\to\H_1$ exists, then
  $\tr_{\H_2}(D)\geq 0$,
\item
  If $E,F\in\L(\H)$ are both positive and $[E,F]=0$, then $EF$ is also
  positive.
\end{myenumerate}
\end{lemma}

\subsection{Representations of compact Lie groups}

In this paragraph, we introduce our notation for finite-dimensional
representations of compact Lie groups. For more details, see, for
example the textbook~\cite{CaSe95} or the introduction
of~\cite{Pf02b}.

Let $G$ be a compact Lie group. We denote finite-dimensional complex
vector spaces on which $G$ is represented by $V_\rho$ and by
$\rho\colon G\to\Aut V_\rho$ the corresponding group
homomorphisms. For each representation $\rho$, the dual representation
is denoted by $\rho^\ast$, and the dual vector space of $V_\rho$ by
$V_\rho^\ast$. The dual representation is given by $\rho^\ast\colon
G\mapsto \Aut V_\rho^\ast$, where
\begin{equation}
\label{eq_dualrep}
  \rho^\ast(g)\colon V_\rho^\ast\to V_\rho^\ast,\quad
    \eta\mapsto \eta\circ\rho(g^{-1}).
\end{equation}
There exists a one-dimensional `trivial' representation of $G$ which
is isomorphic to $\C$.

Since each finite-dimensional representation of $G$ is equivalent to a
unitary representation, we can choose $G$-invariant (sesquilinear)
scalar products $\left<\cdot;\cdot\right>$ and orthonormal bases
$\{e_j\}$. We can then define a bijective antilinear map $\ast\colon
V_\rho\to V_\rho^\ast$ induced by the scalar product,
\begin{equation}
  \ast(v):=(w\mapsto\left<v;w\right>),\qquad v\in V_\rho,
\end{equation}
and construct the dual bases $\{\eta^j\}$ by
$\eta^j:=\ast(e_j)$. Identifying $(V_\rho^\ast)^\ast=V_\rho$, this
yields $\left<e_j;e_k\right>=\eta^j(e_k)=\delta_{jk}$ and furthermore
induces a scalar product on $V_\rho^\ast$, namely
$\bigl<\eta^j;\eta^k\bigr>=\eta^k(e_j)$, $1\leq j,k\leq\dim V_\rho$.

The matrix elements of the representation matrices $\rho(g)$ define
complex valued functions,
\begin{equation}
  t_{jk}^{(\rho)}\colon G\to\C,\qquad g\mapsto t_{jk}^{(\rho)}(g)
  :=\eta^j(\rho(g)e_k)={(\rho(g))}_{jk},
\end{equation}
where $\rho$, $1\leq j,k\leq\dim V_\rho$. They are called
\emph{representative functions} of $G$ and form a commutative and
associative unital algebra over $\C$,
\begin{eqnarray}
  \Calg(G) &:=& \{\,t_{jk}^{(\rho)}\colon\quad
    \rho\mbox{\ finite-dimensional representation of\ }G,\nn\\
  &&\qquad 1\leq j,k\leq\dim V_\rho\,\},
\end{eqnarray}
whose product is given by the matrix elements of the tensor product of
representations,
\begin{eqnarray}
\label{eq_operationprod}
  (t_{jk}^{(\rho)}\cdot t_{\ell m}^{(\sigma)})(g)
  := t_{j\ell,km}^{(\rho\otimes\sigma)}(g),
\end{eqnarray}%
where $1\leq j,k\leq\dim V_\rho$ and $1\leq\ell,m\leq\dim
V_\sigma$. We find the following expressions involving the group unit
$e\in G$,
\begin{equation}
  t_{jk}^{(\rho)}(e) = \delta_{jk},
\end{equation}
products of group elements,
\begin{equation}
\label{eq_copro}
  t_{jk}^{(\rho)}(g\cdot h)=\sum_{\ell=1}^{\dim V_\rho} 
    t_{j\ell}^{(\rho)}(g)\cdot t_{\ell k}^{(\rho)}(h),
\end{equation}
and inverse group elements,
\begin{equation}
  t_{jk}^{(\rho)}(g^{-1})={(\rho(g)^{-1})}_{jk}
    =\overline{{(\rho(g))}_{kj}}=\overline{t_{kj}^{(\rho)}(g)},
\end{equation}
as well as,
\begin{equation}
\label{eq_inverse_dual}
  t_{jk}^{(\rho)}(g^{-1})= \eta^j({\rho(g)}^{-1}e_k) 
    = (\rho^\ast(g)\eta^j)(e_k)
    = \bigl<\eta^k;\rho^\ast(g)\eta^j\bigr> 
    = t_{kj}^{(\rho^\ast)}(g),
\end{equation}
so that for unitary representations, the dual representation is just
the conjugate. The bar denotes complex conjugation.

\subsection{Diagrammatics}
\label{sect_diagrams}

In the following, we introduce the basic object which contains pairs
of intertwiners (or Clebsch--Gordan coefficients) which are together
canonically defined.

\begin{definition}
Let $G$ be a compact Lie group and $\rho_1,\ldots,\rho_r$ be
finite-dimensional irreducible representations of $G$. The \emph{Haar
intertwiner} is defined by
\begin{equation}
\label{eq_haarinter}
  T\colon\bigotimes_{\ell=1}^rV_{\rho_\ell}\to\bigotimes_{\ell=1}^rV_{\rho_\ell},\qquad
   T:=\int_G\rho_1(g)\otimes\cdots\otimes\rho_r(g)\,dg,
\end{equation}%
and has the matrix elements,
\begin{equation}
\label{eq_haarintermatrix}
  T_{m_1m_2\ldots m_r;n_1n_2\ldots n_r} = \int_G
    t^{(\rho_1)}_{m_1n_1}(g)t^{(\rho_2)}_{m_2n_2}(g)\cdots t^{(\rho_r)}_{m_rn_r}(g)\,dg.
\end{equation}
\end{definition}

The following proposition shows how $T$ gives rise to a pair of
intertwiners $P^{(j)}$. It also introduces our normalizations in
detail.

\begin{proposition}
Let $G$ be a compact Lie group and $\rho_1,\ldots,\rho_r$ be
finite-dimensional unitary representations of $G$ such that their
tensor product has the complete decomposition
\begin{equation}
  V_{\rho_1}\otimes\cdots\otimes V_{\rho_r}\cong
  V_{\tau_1}\oplus\cdots\oplus V_{\tau_k},
\end{equation}
into irreducible components $\tau_j$ of which precisely
$\tau_1,\ldots,\tau_\ell$, $0\leq\ell\leq k$, are isomorphic to the
trivial representation. Let $P^{(j)}\colon V_{\rho_1}\otimes\cdots\otimes V_{\rho_r}\to
V_{\tau_j}\subseteq V_{\rho_1}\otimes\cdots\otimes V_{\rho_r}$ be the
$G$-invariant orthogonal projectors associated with the above
decomposition. Then
\begin{equation}
\label{eq_haarinterproj}
  T_{m_1m_2\ldots m_r;n_1n_2\ldots n_r} 
  = \sum_{j=1}^\ell\overline{P^{(j)}_{m_1m_2\ldots m_r}}P^{(j)}_{n_1n_2\ldots n_r},
\end{equation}
where 
\begin{equation}
  P^{(j)}_{n_1n_2\ldots n_r}
  :=\bigl<w^{(j)};e_{n_1}^{(\rho_1)}\otimes e_{n_2}^{(\rho_2)}\otimes\cdots\otimes e_{n_r}^{(\rho_r)}\bigr>,
\end{equation}
are the matrix elements of the projectors. Here $\{e_i^{(\rho_q)}\}$
denotes an orthonormal basis of $V_{\rho_q}$ and $w^{(j)}$ a
normalized vector spanning $V_{\tau_j}\subseteq
V_{\rho_1}\otimes\cdots\otimes V_{\rho_r}$.
\end{proposition}

\begin{figure}[t]
\begin{center}
\input{pstex/diagrams.pstex_t}
\end{center}
\mycaption{fig_diagrams}{%
The basic diagrams in order to describe representations and morphisms
of $G$. Refer to Section~\ref{sect_diagrams} for detailed explanations.}
\end{figure}

Equation~\eqref{eq_haarinterproj} shows how the canonical object $T$
is decomposed into pairs of intertwiners $P^{(j)}$. In the
Ponzano--Regge model, for example, the symmetry group is $G=\SU(2)$,
and the assignment of $6j$-symbols to the tetrahedra can be obtained
as a special case of the dual formulation of non-Abelian lattice gauge
theory~\cite{OePf01,Pf01,Oe02}. For each triangle we have a Haar
intertwiner $T\colon V_j\otimes V_k\otimes V_\ell\to V_j\otimes
V_k\otimes V_\ell$. The projectors $P^{(j)}$
in~\eqref{eq_haarinterproj} are then $\SU(2)$ intertwiners as
in~\eqref{eq_su2trivalent} and belong to two different $6j$-symbols
associated with the two tetrahedra attached to the triangle.

In order to perform calculations involving the Haar intertwiner, there
exists a convenient diagrammatic language which can be understood as a
specialization of the Reshetikhin--Turaev ribbon
diagrams~\cite{ReTu90} to representations of compact Lie groups.

\begin{figure}[t]
\begin{center}
\input{pstex/gaugefix.pstex_t}
\end{center}
\mycaption{fig_gaugefix}{%
The gauge fixing relation, Proposition~\ref{prop_haar}(5).}
\end{figure}

\begin{figure}[t]
\begin{center}
\input{pstex/ind_start.pstex_t}
\end{center}
\mycaption{fig_ind_start}{%
The anchor of the induction proof, see Lemma~\ref{lemma_start}.}
\end{figure}

Figure~\ref{fig_diagrams} shows the basic diagrams. These are read
from top to bottom. We draw directed lines which are labeled with
finite-dimensional unitary representations $\rho$ of $G$. If the arrow
points down, the line denotes the identity map of $V_\rho$,
Figure~\ref{fig_diagrams}(a). If the arrow points up as in~(b), it
refers to the identity map of the dual representation
$V_\rho^\ast$. Placing symbols next to each other corresponds to the
tensor product, placing symbols below each other denotes the
composition of maps. The diagrams~(c) and~(d) show co-evaluation and
evaluation,
\begin{alignat}{2}
  \coev_{\rho}&\colon\C\to V_\rho\otimes V_\rho^\ast,&\qquad&
    1\mapsto\sum_{j=1}^{\dim V_\rho}v_j\otimes\eta^j,\\
  \ev_{\rho}&\colon V^\ast_\rho\otimes V_\rho\to\C,&\qquad&
    \alpha\otimes w\mapsto \alpha(w).
\end{alignat}
Two tensor factors are swapped by the map $\psi_{\rho,\sigma}\colon
V_\rho\otimes V_\sigma\to V_\sigma\otimes V_\rho, v\otimes w\mapsto
w\otimes v$ in diagram~(e). The Haar
intertwiner~\eqref{eq_haarintermatrix} is shown in~(f). The trivial
representation is invisible in these diagrams. Note that for
representations of groups, as opposed to quantum groups or super
groups, our diagrams do not involve any framing nor any non-trivial
braiding. Any diagram that can be written down using only the symbols
of Figure~\ref{fig_diagrams}, is $G$-covariant, \ie\ represents a
$G$-morphism. The Haar intertwiner satisfies special
properties~\cite{Oe02} which are summarized by the following
proposition.

\begin{proposition}
\label{prop_haar}
Let $G$ be a compact Lie group and $T$ denote the Haar
intertwiner~\eqref{eq_haarinter} for finite-dimensional unitary
representations of $G$.
\begin{myenumerate}
\item
  $T$ is a $G$-morphism,
\item
  $T^2=T$,
\item
  $T^\dagger=T$,
\item
  If $\Phi$ is a $G$-morphism, then $\Phi\circ T=T\circ\Phi$,
\item
  $T$ satisfies the gauge fixing relation, \ie\ in any diagram in
  which we can draw a closed loop (the dashed line in
  Figure~\ref{fig_gaugefix}(a)) which intersects only the boxes of
  Haar intertwiners, but no representation lines, then we may replace
  one of the Haar intertwiners by the identity morphism
  (Figure~\ref{fig_gaugefix}(b)). The step of going from~(b) to~(a) is
  called \emph{inverse gauge fixing}.
\end{myenumerate}
\end{proposition}

\begin{definition}
Let $\rho_1,\ldots,\rho_k$ be finite-dimensional unitary
representations of $G$. A diagram which represents a morphism
$\Phi\colon V_{\rho_1}\otimes\cdots\otimes V_{\rho_k}\to
V_{\rho_1}\otimes\cdots\otimes V_{\rho_k}$, is called \emph{positive}
if $\Phi$ is a positive linear operator for any choice of
finite-dimensional unitary representations for the
$\rho_1,\ldots,\rho_k$ and for the internal lines of the diagram.
\end{definition}

If we talk about positive diagrams, we can obviously omit the arrows
from the diagrams because the above definition states a condition on
all representations including in particular the dual ones. Our basic
example of a positive diagram is the Haar intertwiner,
Figure~\ref{fig_diagrams}(f), or any partial trace of it.

%
\section{Positivity of relativistic spin network\\ evaluations}
%
\label{sect_positive}

\subsection{Positivity proof}

\begin{figure}[t]
\begin{center}
\input{pstex/ind_claim.pstex_t}
\end{center}
\mycaption{fig_ind_claim}{%
Theorem~\ref{thm_induction} establishes
that this diagram is positive.  It consists of $k$ Haar intertwiners
with $k+1$ lines each. For each such Haar intertwiner, one line is an
external one and one is a closed loop. Additionally, for each pair of
Haar intertwiners, there is one loop linking the two.}
\end{figure}

\begin{figure}[t]
\begin{center}
\input{pstex/ind_step.pstex_t}
\end{center}
\mycaption{fig_ind_step}{%
Diagrams used in the proof of Theorem~\ref{thm_induction}.}
\end{figure}

In this section, we generate an infinite family of positive diagrams by
induction. The anchor of the induction is the following lemma.

\begin{lemma}
\label{lemma_start}
The diagram in Figure~\ref{fig_ind_start}(e) is positive. 
\end{lemma}

\begin{proof}
We start with the diagrams in Figure~\ref{fig_ind_start}(a)
and~(b). Diagram~(a) is positive as a partial trace of the Haar
intertwiner. Since it denotes a representation morphism, it commutes
with the Haar intertwiner~(b) by
Proposition~\ref{prop_haar}(4). Lemma~\ref{lemma_positivity}(4) then
implies the positivity of~(c). From there we obtain~(d) by a sequence
of inverse gauge fixing and gauge fixing along the dashed line
in~(c). Finally, (e) is positive as a partial trace of~(d).
\end{proof}

We wish to generalize this result to the diagram that generalizes
Figure~\ref{fig_ind_start}(e) to $k$ Haar intertwiners and which is
shown in Figure~\ref{fig_ind_claim}. This diagram consists of $k$ Haar
intertwiners with $k+1$ lines each. For each such Haar intertwiner,
one line is an external one and one is a closed loop. Additionally,
for each pair of Haar intertwiners, there is one loop linking the two.

\begin{theorem}
\label{thm_induction}
The diagram of Figure~\ref{fig_ind_claim} is positive for any number
$k$ of Haar intertwiners.
\end{theorem}

\begin{proof}
We assume that the theorem is true for $k-1$ (for $k-1=2$, this was
proved in Lemma~\ref{lemma_start}). In order to keep the drawings
simple, Figure~\ref{fig_ind_step} shows the case $k=3$. The argument
is, of course, independent of $k$.

The diagram in Figure~\ref{fig_ind_step}(a) is positive by
assumption. Note that we are allowed to replace single lines by double
lines in any positive diagram because positivity holds for any
assignment of representations, in particular for tensor
products. Diagram~(a) denotes a morphism so that it commutes with the
Haar intertwiner with four or more lines. Therefore diagram~(b) is
positive by Lemma~\ref{lemma_positivity}(4). We obtain~(c) by inverse
gauge fixing and gauge fixing along the dashed line in~(b). The
proposition follows by exchanging tensor factors and taking partial
traces.
\end{proof}

\subsection{Relativistic spin network evaluations}

\begin{figure}[t]
\begin{center}
\input{pstex/graph.pstex_t}
\end{center}
\mycaption{fig_graph}{%
The structure of the positive diagrams~(a) generated by
Theorem~\ref{thm_induction} is that of a complete graph~(b) where the
representations are of the form $V\otimes V^\ast$ and the intertwiner
is given by a group integration~(c).}
\end{figure}

We have shown in Theorem~\ref{thm_induction} that the diagram of
Figure~\ref{fig_ind_claim} is positive. Now we choose the trivial
representation for all external lines and for the little loops that
are attached only to one Haar intertwiner, and obtain positive
diagrams such as that in Figure~\ref{fig_graph}(a) which was drawn for
$k=4$. If we view each pair of lines that belong to the same loop as a
representation $V\otimes V^\ast$, these diagrams have the structure of
the complete graph of $k$ vertices, Figure~\ref{fig_graph}(b). The
spin network of Figure~\ref{fig_graph}(a) is a relativistic spin
network in the following sense.

\begin{definition}
\label{def_relativistic}
Let $G$ be a compact Lie group. A \emph{relativistic spin network}
with symmetry $G\times G$ is a spin network whose edges are labeled
with representations of the form $V\otimes V^\ast$ of $G\times G$
where $V$ denotes a representation of $G$, and whose vertices are
labeled by the intertwiner of Figure~\ref{fig_graph}(c), given by one
integration over $G$.
\end{definition}

\begin{remark}
\begin{myenumerate}
\item
  The relativistic spin networks used in the model of
  Barrett--Crane~\cite{BaCr98} form a special case of
  Definition~\ref{def_relativistic} for $G=\SU(2)$, using
  $\Spin(4)\cong\SU(2)\times\SU(2)$. For $\SU(2)$, the balanced
  representations $V\otimes V$ are isomorphic to our representations
  $V\otimes V^\ast$, and Figure~\ref{fig_graph}(c) is precisely the
  presentation~\cite{Ba98} of the Barrett--Crane intertwiner as an
  integral over $\SU(2)\cong S^3$.
\item
  Observe that the choice of $V\otimes V^\ast$ instead of the
  isomorphic $V\otimes V$ has eliminated a number of signs as we have
  already observed in~\cite{Pf02a}. There we have also shown that the
  choice $V\otimes V^\ast$ is the canonical one compatible with the
  integral presentation of the Barrett--Crane intertwiner.
\item
  Balanced representations of $\Spin(4)$ factor through the covering
  map $\Spin(4)\to\SO(4)$ and therefore form representations of
  $\SO(4)$.
\end{myenumerate}
\end{remark}

Reisenberger~\cite{Re99} has proved that the space of Barrett--Crane
intertwiners for any given valence of the vertex is
one-dimensional. The particular prefactor can depend on the
conventions used.  For our definition using the integral presentation
of the Barrett--Crane intertwiner and balanced representations of the
form $V\otimes V^\ast$, the relativistic spin networks evaluate to
non-negative numbers. If one uses, however, Kauffman's
conventions~\cite{KaLi94} for $\SU(2)$ spin networks, there is an
additional overall sign which depends on the representations. This is
the sign~\cite{BaCh02} that cancels only if one multiplies all spin
networks associated to the four-simplices of a closed manifold. This
sign is not essential to the spin network evaluation, but rather an
artifact of conventions which are unnatural in the present context.

Theorem~\ref{thm_induction} has shown that relativistic spin networks
on the complete graph of $k$ vertices evaluate to non-negative
numbers. This result can be specialized to any subgraph of the
complete graph by choosing the trivial representation for all edges
that are missing compared to the complete graph. We have therefore
proved

\begin{corollary}
Let $G$ be a compact Lie group. Any relativistic spin network with
symmetry $G\times G$ on any subgraph of a complete graph evaluates to
a non-negative real number.
\end{corollary}

The positivity of relativistic spin network evaluations proved so far
implies that the individual summands of the partition function of the
Barrett--Crane model on a generic two-complex are non-negative because
these summands are products of various amplitudes each of which is
calculated by evaluating relativistic spin networks. This result can
be extended to prove the absence of destructive interference in the
Riemannian Barrett--Crane model on any two-complex, using Corollary~1
of~\cite{BaCh02}.

%
\section{Counterexamples to the stronger conjecture}
%
\label{sect_counterexample}

\begin{figure}[t]
\begin{center}
\input{pstex/counterexample.pstex_t}
\end{center}
\mycaption{fig_counterexample}{%
These diagrams are not positive.}
\end{figure}

\begin{table}
\begin{center}
\begin{tabular}{l|rrr}
  $S_3$            & $[1^+]$ &  $[2]$ & $[1^-]$ \\
\hline
  $()$             & $1$   &  $2$ &  $1$  \\
  $(12),(13),(23)$ & $1$   &  $0$ & $-1$  \\ 
  $(123),(132)$    & $1$   & $-1$ &  $1$  \\ 
\end{tabular}
\end{center}
\caption{\label{tab_character} The character table of the finite group
$S_3$. The rows are labeled by its conjugacy classes and the columns
by the finite-dimensional irreducible representations.}
\end{table}

Given the result of Section~\ref{sect_positive}, it is tempting to
conjecture that the use of the Haar intertwiner is the magical
ingredient that renders all these spin networks non-negative. However,
the stronger conjecture that all diagrams are positive if they are
composed only from the building blocks shown in
Figure~\ref{fig_diagrams}, is not true.

As a first counterexample we consider the diagram in
Figure~\ref{fig_counterexample}(a). The compact Lie group $G$ is the
finite group $S_3$ (with the discrete topology). Its character table
is given in Table~\ref{tab_character}. Let $[2]$ denote the
two-dimensional representation of $S_3$ and $[1^-]$ the
one-dimensional parity representation. If we choose
$\rho_1=\rho_2=\rho_3=[2]$ and $\rho_4=[1^-]$, a direct calculation
shows that Figure~\ref{fig_counterexample}(a) evaluates to
\begin{equation}
  \frac{1}{{|S_3|}^3}\sum_{f,g,h\in S_3}\chi^{[2]}(fg)\chi^{[2]}(fh)\chi^{[2]}(gh)\chi^{[1^-]}(h)=-\frac{1}{4}.
\end{equation}

\begin{figure}[t]
\begin{center}
\input{pstex/simplex.pstex_t}
\end{center}
\mycaption{fig_simplex}{%
The diagram appearing in lattice gauge theory on the two-complex dual
to a triangulation of the sphere $S^3$ by two tetrahedra.}
\end{figure}

The stronger conjecture is therefore false, at least as long as we do
not restrict the class of allowed Lie groups or the class of diagrams.

The diagrams studied in Section~\ref{sect_positive} are obviously
special in the sense that they are related to relativistic spin
networks. Is the counterexample presented above maybe too pathologic?
It is instructive to re-arrange the positive diagrams of
Section~\ref{sect_positive}. For $k=4$, we have
Figure~\ref{fig_graph}(a) which can be drawn as
Figure~\ref{fig_simplex}. This is the diagram which appears in the
study of lattice gauge theory on the two-complex dual to a
triangulation of the sphere $S^3$ by two tetrahedra. One of the
tetrahedra is located at the center of the diagram, the other one at
infinity. In the language of~\cite{Oe02}, the diagram is the circuit
diagram in the two-complex dual to the cellular decomposition defined
by the triangulation. The diagram for general $k$ is the circuit
diagram for the triangulation of $S^{k-1}$ by two $(k-1)$-simplices.

Is it maybe true that the circuit diagrams of all triangulations or of
all cellular decompositions of $S^{k-1}$ have a positive evaluation?
The answer is again negative as our second counterexample shows.

Consider the diagram in Figure~\ref{fig_simplex}. Subdivide the
central tetrahedron into four tetrahedra ($1\leftrightarrow 4$ Pachner
move) and draw the circuit diagram for this finer triangulation. By
gauge fixing and substituting the trivial representation for some
lines, we arrive at the diagram of
Figure~\ref{fig_counterexample}(b). This diagram evaluates to a
negative number for some choice of representations which implies that
the diagram of the refined triangulation of $S^3$ is negative for some
labeling.

In order to see this, consider Figure~\ref{fig_counterexample}(b) and
gauge fix again, removing the Haar intertwiner marked by a
`$\ast$'. For any assignment of irreducible representations to the
lines, a number of Haar intertwiners are trivial and can be explicitly
evaluated. We remove two of them, marked by `-' from the diagram. The
resulting diagram can be computed by hand for $G=S_3$. We choose all
representations to be $[2]$ except for the line indicated in
Figure~\ref{fig_counterexample}(b) with is labeled by $[1^-]$. The
calculation is completely analogous to our first counterexample, and
the diagram evaluates to $-1/8$.

We have therefore shown that the stronger conjecture fails even if one
restricts the class of diagrams to circuit diagrams of triangulations
of the sphere $S^3$.

%
\section{Discussion}
%
\label{sect_conclusion}

First we point out a general difficulty with the definition of the
Barrett--Crane model. Its vertex amplitude is found by geometrical
conditions to be the `relativistic $10j$-symbol', defined by the
requirement that its representations are balanced and that its
intertwiners are Barrett--Crane intertwiners. This intertwiner is
\emph{a priori} only specified up to a complex factor. It would
obviously be a disaster if, as a consequence, the full $10j$-symbol
contains an arbitrary complex factor. The standard strategy to avoid
such an ambiguity is to fix the conventions for all intertwiners
throughout the representation category of $\SU(2)$ in a systematic
way, for example, as in~\cite{KaLi94}. The remaining ambiguity is then
still a representation dependent sign.

The construction presented here is, on the contrary, completely
canonical. There are no arbitrary signs and we are in addition
rewarded by a special property, namely the positivity of any single
relativistic spin network. It should be pointed out that the
Lorentzian versions of the relativistic spin networks were defined in
terms of their integral presentation right in the
beginning~\cite{BaCr00}. This definition is canonical, and there are
no similar sign ambiguities there.

We observe that the framework developed in the present article extends
to non-compact Lie groups $G$ provided their representations are
unitary and that one can show the existence of all relevant
traces. The Lorentzian version of relativistic spin
networks~\cite{BaCr00}, however, has a different structure and is not
covered by our result.

What are possible application of our positivity result? The
Riemannian, \ie\ $\Spin(4)$- or $\SO(4)$-symmetric, Barrett--Crane
model can be defined on any two-complex, not just on a two-complex
dual to a triangulation~\cite{Ba98a}, leading to relativistic spin
networks on general graphs as the vertex amplitudes rather than just
$10j$-symbols. Our result provides a canonical definition of these
spin networks and establishes the positivity of each single
diagram. As a consequence, one can show the absence of destructive
interference following~\cite{BaCh02}. This result supports the
conjecture that the Barrett--Crane model does not define any unitary
evolution operator, but rather some projector for which the spin
network basis is very special and gives rise to only positive (or only
negative) matrix elements.

In contrast to the Barrett--Crane model, lattice gauge theory and
lattice sigma models are meant to be models of Statistical Mechanics
with positive weights that admit a probability interpretation. For
non-Abelian lattice gauge theory, our counterexample shows that the
strong-weak dual spin foam model~\cite{OePf01, Pf01,Oe02} does not in
general have positive amplitudes. In order to apply Monte Carlo
techniques, one therefore needs a special treatment of the signs. The
situation for the spin network models strong-weak dual to lattice
sigma models~\cite{Pf02b} is much better. Both the $G\times
G$-symmetric lattice chiral model and the $\SO(4)$-symmetric lattice
non-linear sigma model, also called the $3$-vector model, have dual
descriptions in terms of relativistic spin networks so that we have
non-negative amplitudes in these cases.

\acknowledgments

I would like to thank Robert Oeckl for stimulating discussions and for
comments on the manuscript.

\end{document}

%% file: pstex/diagrams.pstex_t
\begin{picture}(0,0)%
\includegraphics{pstex/diagrams.pstex}%
\end{picture}%
\setlength{\unitlength}{3947sp}%
\begingroup\makeatletter\ifx\SetFigFont\undefined%
\gdef\SetFigFont#1#2#3#4#5{%
  \reset@font\fontsize{#1}{#2pt}%
  \fontfamily{#3}\fontseries{#4}\fontshape{#5}%
  \selectfont}%
\fi\endgroup%
\begin{picture}(5787,2265)(1576,-2161)
\put(6601,-811){\makebox(0,0)[lb]{\smash{\SetFigFont{12}{14.4}{\rmdefault}{\mddefault}{\updefault}{\color[rgb]{0,0,0}$T$}%
}}}
\put(6976,-361){\makebox(0,0)[b]{\smash{\SetFigFont{10}{12.0}{\rmdefault}{\mddefault}{\updefault}{\color[rgb]{0,0,0}$\cdots$}%
}}}
\put(6976,-1486){\makebox(0,0)[b]{\smash{\SetFigFont{10}{12.0}{\rmdefault}{\mddefault}{\updefault}{\color[rgb]{0,0,0}$\cdots$}%
}}}
\put(6601,-61){\makebox(0,0)[b]{\smash{\SetFigFont{10}{12.0}{\rmdefault}{\mddefault}{\updefault}{\color[rgb]{0,0,0}$n_1$}%
}}}
\put(7201,-61){\makebox(0,0)[b]{\smash{\SetFigFont{10}{12.0}{\rmdefault}{\mddefault}{\updefault}{\color[rgb]{0,0,0}$n_r$}%
}}}
\put(7201,-1861){\makebox(0,0)[b]{\smash{\SetFigFont{10}{12.0}{\rmdefault}{\mddefault}{\updefault}{\color[rgb]{0,0,0}$m_r$}%
}}}
\put(6601,-1861){\makebox(0,0)[b]{\smash{\SetFigFont{10}{12.0}{\rmdefault}{\mddefault}{\updefault}{\color[rgb]{0,0,0}$m_1$}%
}}}
\put(2476,-2161){\makebox(0,0)[b]{\smash{\SetFigFont{12}{14.4}{\rmdefault}{\mddefault}{\updefault}{\color[rgb]{0,0,0}$(b)$}%
}}}
\put(2476,-886){\makebox(0,0)[b]{\smash{\SetFigFont{14}{16.8}{\rmdefault}{\mddefault}{\updefault}{\color[rgb]{0,0,0}$=$}%
}}}
\put(3526,-2161){\makebox(0,0)[b]{\smash{\SetFigFont{12}{14.4}{\rmdefault}{\mddefault}{\updefault}{\color[rgb]{0,0,0}$(c)$}%
}}}
\put(4576,-2161){\makebox(0,0)[b]{\smash{\SetFigFont{12}{14.4}{\rmdefault}{\mddefault}{\updefault}{\color[rgb]{0,0,0}$(d)$}%
}}}
\put(1576,-361){\makebox(0,0)[rb]{\smash{\SetFigFont{10}{12.0}{\rmdefault}{\mddefault}{\updefault}{\color[rgb]{0,0,0}$\rho$}%
}}}
\put(2176,-361){\makebox(0,0)[rb]{\smash{\SetFigFont{10}{12.0}{\rmdefault}{\mddefault}{\updefault}{\color[rgb]{0,0,0}$\rho$}%
}}}
\put(2776,-361){\makebox(0,0)[lb]{\smash{\SetFigFont{10}{12.0}{\rmdefault}{\mddefault}{\updefault}{\color[rgb]{0,0,0}$\rho^\ast$}%
}}}
\put(3226,-1411){\makebox(0,0)[rb]{\smash{\SetFigFont{10}{12.0}{\rmdefault}{\mddefault}{\updefault}{\color[rgb]{0,0,0}$\rho$}%
}}}
\put(4276,-361){\makebox(0,0)[rb]{\smash{\SetFigFont{10}{12.0}{\rmdefault}{\mddefault}{\updefault}{\color[rgb]{0,0,0}$\rho$}%
}}}
\put(6901,-2161){\makebox(0,0)[b]{\smash{\SetFigFont{12}{14.4}{\rmdefault}{\mddefault}{\updefault}{\color[rgb]{0,0,0}$(f)$}%
}}}
\put(5626,-2161){\makebox(0,0)[b]{\smash{\SetFigFont{12}{14.4}{\rmdefault}{\mddefault}{\updefault}{\color[rgb]{0,0,0}$(e)$}%
}}}
\put(5926,-361){\makebox(0,0)[lb]{\smash{\SetFigFont{10}{12.0}{\rmdefault}{\mddefault}{\updefault}{\color[rgb]{0,0,0}$\sigma$}%
}}}
\put(5476,-361){\makebox(0,0)[lb]{\smash{\SetFigFont{10}{12.0}{\rmdefault}{\mddefault}{\updefault}{\color[rgb]{0,0,0}$\rho$}%
}}}
\put(1651,-2161){\makebox(0,0)[b]{\smash{\SetFigFont{12}{14.4}{\rmdefault}{\mddefault}{\updefault}{\color[rgb]{0,0,0}$(a)$}%
}}}
\end{picture}

%% file: pstex/gaugefix.pstex_t
\begin{picture}(0,0)%
\includegraphics{pstex/gaugefix.pstex}%
\end{picture}%
\setlength{\unitlength}{3947sp}%
\begingroup\makeatletter\ifx\SetFigFont\undefined%
\gdef\SetFigFont#1#2#3#4#5{%
  \reset@font\fontsize{#1}{#2pt}%
  \fontfamily{#3}\fontseries{#4}\fontshape{#5}%
  \selectfont}%
\fi\endgroup%
\begin{picture}(5874,2712)(364,-2611)
\put(1876,-2611){\makebox(0,0)[b]{\smash{\SetFigFont{12}{14.4}{\rmdefault}{\mddefault}{\updefault}{\color[rgb]{0,0,0}$(a)$}%
}}}
\put(4876,-2611){\makebox(0,0)[b]{\smash{\SetFigFont{12}{14.4}{\rmdefault}{\mddefault}{\updefault}{\color[rgb]{0,0,0}$(b)$}%
}}}
\put(3451,-1111){\makebox(0,0)[b]{\smash{\SetFigFont{14}{16.8}{\rmdefault}{\mddefault}{\updefault}{\color[rgb]{0,0,0}$=$}%
}}}
\end{picture}

%% file: pstex/ind_start.pstex_t
\begin{picture}(0,0)%
\includegraphics{pstex/ind_start.pstex}%
\end{picture}%
\setlength{\unitlength}{3947sp}%
\begingroup\makeatletter\ifx\SetFigFont\undefined%
\gdef\SetFigFont#1#2#3#4#5{%
  \reset@font\fontsize{#1}{#2pt}%
  \fontfamily{#3}\fontseries{#4}\fontshape{#5}%
  \selectfont}%
\fi\endgroup%
\begin{picture}(5724,2337)(1714,-2011)
\put(2851,-2011){\makebox(0,0)[b]{\smash{\SetFigFont{12}{14.4}{\rmdefault}{\mddefault}{\updefault}{\color[rgb]{0,0,0}$(b)$}%
}}}
\put(2101,-2011){\makebox(0,0)[b]{\smash{\SetFigFont{12}{14.4}{\rmdefault}{\mddefault}{\updefault}{\color[rgb]{0,0,0}$(a)$}%
}}}
\put(6826,-2011){\makebox(0,0)[b]{\smash{\SetFigFont{12}{14.4}{\rmdefault}{\mddefault}{\updefault}{\color[rgb]{0,0,0}$(e)$}%
}}}
\put(3901,-2011){\makebox(0,0)[b]{\smash{\SetFigFont{12}{14.4}{\rmdefault}{\mddefault}{\updefault}{\color[rgb]{0,0,0}$(c)$}%
}}}
\put(5476,-2011){\makebox(0,0)[b]{\smash{\SetFigFont{12}{14.4}{\rmdefault}{\mddefault}{\updefault}{\color[rgb]{0,0,0}$(d)$}%
}}}
\put(4726,-736){\makebox(0,0)[b]{\smash{\SetFigFont{14}{16.8}{\rmdefault}{\mddefault}{\updefault}{\color[rgb]{0,0,0}$=$}%
}}}
\end{picture}

%% file: pstex/ind_claim.pstex_t
\begin{picture}(0,0)%
\includegraphics{pstex/ind_claim.pstex}%
\end{picture}%
\setlength{\unitlength}{3947sp}%
\begingroup\makeatletter\ifx\SetFigFont\undefined%
\gdef\SetFigFont#1#2#3#4#5{%
  \reset@font\fontsize{#1}{#2pt}%
  \fontfamily{#3}\fontseries{#4}\fontshape{#5}%
  \selectfont}%
\fi\endgroup%
\begin{picture}(4149,2124)(1189,-1723)
\put(4201,-736){\makebox(0,0)[b]{\smash{\SetFigFont{12}{14.4}{\rmdefault}{\mddefault}{\updefault}{\color[rgb]{0,0,0}$\ldots$}%
}}}
\put(5101,-136){\makebox(0,0)[b]{\smash{\SetFigFont{10}{12.0}{\rmdefault}{\mddefault}{\updefault}{\color[rgb]{0,0,0}$\ldots$}%
}}}
\put(5101,-1261){\makebox(0,0)[b]{\smash{\SetFigFont{10}{12.0}{\rmdefault}{\mddefault}{\updefault}{\color[rgb]{0,0,0}$\ldots$}%
}}}
\put(1651,-136){\makebox(0,0)[b]{\smash{\SetFigFont{10}{12.0}{\rmdefault}{\mddefault}{\updefault}{\color[rgb]{0,0,0}$\ldots$}%
}}}
\put(1651,-1261){\makebox(0,0)[b]{\smash{\SetFigFont{10}{12.0}{\rmdefault}{\mddefault}{\updefault}{\color[rgb]{0,0,0}$\ldots$}%
}}}
\put(2626,-136){\makebox(0,0)[b]{\smash{\SetFigFont{10}{12.0}{\rmdefault}{\mddefault}{\updefault}{\color[rgb]{0,0,0}$\ldots$}%
}}}
\put(2626,-1261){\makebox(0,0)[b]{\smash{\SetFigFont{10}{12.0}{\rmdefault}{\mddefault}{\updefault}{\color[rgb]{0,0,0}$\ldots$}%
}}}
\put(3601,-1261){\makebox(0,0)[b]{\smash{\SetFigFont{10}{12.0}{\rmdefault}{\mddefault}{\updefault}{\color[rgb]{0,0,0}$\ldots$}%
}}}
\put(3601,-136){\makebox(0,0)[b]{\smash{\SetFigFont{10}{12.0}{\rmdefault}{\mddefault}{\updefault}{\color[rgb]{0,0,0}$\ldots$}%
}}}
\end{picture}

%% file: pstex/ind_step.pstex_t
\begin{picture}(0,0)%
\includegraphics{pstex/ind_step.pstex}%
\end{picture}%
\setlength{\unitlength}{3947sp}%
\begingroup\makeatletter\ifx\SetFigFont\undefined%
\gdef\SetFigFont#1#2#3#4#5{%
  \reset@font\fontsize{#1}{#2pt}%
  \fontfamily{#3}\fontseries{#4}\fontshape{#5}%
  \selectfont}%
\fi\endgroup%
\begin{picture}(5574,3162)(1789,-2611)
\put(4426,-2611){\makebox(0,0)[b]{\smash{\SetFigFont{12}{14.4}{\rmdefault}{\mddefault}{\updefault}{\color[rgb]{0,0,0}$(b)$}%
}}}
\put(5626,-886){\makebox(0,0)[b]{\smash{\SetFigFont{14}{16.8}{\rmdefault}{\mddefault}{\updefault}{\color[rgb]{0,0,0}$=$}%
}}}
\put(6601,-2611){\makebox(0,0)[b]{\smash{\SetFigFont{12}{14.4}{\rmdefault}{\mddefault}{\updefault}{\color[rgb]{0,0,0}$(c)$}%
}}}
\put(2476,-2611){\makebox(0,0)[b]{\smash{\SetFigFont{12}{14.4}{\rmdefault}{\mddefault}{\updefault}{\color[rgb]{0,0,0}$(a)$}%
}}}
\end{picture}

%% file: pstex/graph.pstex_t
\begin{picture}(0,0)%
\includegraphics{pstex/graph.pstex}%
\end{picture}%
\setlength{\unitlength}{3947sp}%
\begingroup\makeatletter\ifx\SetFigFont\undefined%
\gdef\SetFigFont#1#2#3#4#5{%
  \reset@font\fontsize{#1}{#2pt}%
  \fontfamily{#3}\fontseries{#4}\fontshape{#5}%
  \selectfont}%
\fi\endgroup%
\begin{picture}(5949,3234)(1039,-2611)
\put(4201,-2611){\makebox(0,0)[b]{\smash{\SetFigFont{12}{14.4}{\rmdefault}{\mddefault}{\updefault}{\color[rgb]{0,0,0}$(b)$}%
}}}
\put(2251,-2611){\makebox(0,0)[b]{\smash{\SetFigFont{12}{14.4}{\rmdefault}{\mddefault}{\updefault}{\color[rgb]{0,0,0}$(a)$}%
}}}
\put(6301,-2611){\makebox(0,0)[b]{\smash{\SetFigFont{12}{14.4}{\rmdefault}{\mddefault}{\updefault}{\color[rgb]{0,0,0}$(c)$}%
}}}
\put(5551,-961){\makebox(0,0)[b]{\smash{\SetFigFont{12}{14.4}{\rmdefault}{\mddefault}{\updefault}{\color[rgb]{0,0,0}$=$}%
}}}
\end{picture}

%% file: pstex/counterexample.pstex_t
\begin{picture}(0,0)%
\includegraphics{pstex/counterexample.pstex}%
\end{picture}%
\setlength{\unitlength}{3947sp}%
\begingroup\makeatletter\ifx\SetFigFont\undefined%
\gdef\SetFigFont#1#2#3#4#5{%
  \reset@font\fontsize{#1}{#2pt}%
  \fontfamily{#3}\fontseries{#4}\fontshape{#5}%
  \selectfont}%
\fi\endgroup%
\begin{picture}(5740,3898)(1309,-3211)
\put(1876,-211){\makebox(0,0)[b]{\smash{\SetFigFont{12}{14.4}{\rmdefault}{\mddefault}{\updefault}{\color[rgb]{0,0,0}$1$}%
}}}
\put(1726,-961){\makebox(0,0)[b]{\smash{\SetFigFont{12}{14.4}{\rmdefault}{\mddefault}{\updefault}{\color[rgb]{0,0,0}$2$}%
}}}
\put(1801,389){\makebox(0,0)[rb]{\smash{\SetFigFont{10}{12.0}{\rmdefault}{\mddefault}{\updefault}{\color[rgb]{0,0,0}$\rho_1$}%
}}}
\put(1351,-511){\makebox(0,0)[rb]{\smash{\SetFigFont{10}{12.0}{\rmdefault}{\mddefault}{\updefault}{\color[rgb]{0,0,0}$\rho_3$}%
}}}
\put(2326, 89){\makebox(0,0)[lb]{\smash{\SetFigFont{10}{12.0}{\rmdefault}{\mddefault}{\updefault}{\color[rgb]{0,0,0}$\rho_2$}%
}}}
\put(2026,-1411){\makebox(0,0)[lb]{\smash{\SetFigFont{10}{12.0}{\rmdefault}{\mddefault}{\updefault}{\color[rgb]{0,0,0}$\rho_4$}%
}}}
\put(5176,-3211){\makebox(0,0)[b]{\smash{\SetFigFont{12}{14.4}{\rmdefault}{\mddefault}{\updefault}{\color[rgb]{0,0,0}$(b)$}%
}}}
\put(1876,-3211){\makebox(0,0)[b]{\smash{\SetFigFont{12}{14.4}{\rmdefault}{\mddefault}{\updefault}{\color[rgb]{0,0,0}$(a)$}%
}}}
\put(1726,-1861){\makebox(0,0)[b]{\smash{\SetFigFont{12}{14.4}{\rmdefault}{\mddefault}{\updefault}{\color[rgb]{0,0,0}$3$}%
}}}
\put(3826,-1336){\makebox(0,0)[b]{\smash{\SetFigFont{12}{14.4}{\rmdefault}{\mddefault}{\updefault}{\color[rgb]{0,0,0}$\ast$}%
}}}
\put(4201,-361){\makebox(0,0)[b]{\smash{\SetFigFont{12}{14.4}{\rmdefault}{\mddefault}{\updefault}{\color[rgb]{0,0,0}$-$}%
}}}
\put(5176,-2311){\makebox(0,0)[b]{\smash{\SetFigFont{12}{14.4}{\rmdefault}{\mddefault}{\updefault}{\color[rgb]{0,0,0}$-$}%
}}}
\put(4876,-1486){\makebox(0,0)[lb]{\smash{\SetFigFont{10}{12.0}{\rmdefault}{\mddefault}{\updefault}{\color[rgb]{0,0,0}$[1^-]$}%
}}}
\end{picture}

%% file: pstex/simplex.pstex_t
\begin{picture}(0,0)%
\includegraphics{pstex/simplex.pstex}%
\end{picture}%
\setlength{\unitlength}{3947sp}%
\begingroup\makeatletter\ifx\SetFigFont\undefined%
\gdef\SetFigFont#1#2#3#4#5{%
  \reset@font\fontsize{#1}{#2pt}%
  \fontfamily{#3}\fontseries{#4}\fontshape{#5}%
  \selectfont}%
\fi\endgroup%
\begin{picture}(2932,2912)(601,-2224)
\end{picture}